\documentstyle[psfig]{mnnd}

\newcommand{\be}{\begin{equation}}
\newcommand{\ee}{\end{equation}}
\newcommand{\ba}{\begin{array}}
\newcommand{\ea}{\end{array}}
\newcommand{\bea}{\begin{eqnarray}}
\newcommand{\eea}{\end{eqnarray}}

\newcommand{\mbf}[1]{{\bf #1}}

\newcommand{\DS}{\displaystyle}

\newcommand{\Dot}[1]{\stackrel{\mbox{.}}{#1}}
\newcommand{\DDot}[1]{\stackrel{\mbox{..}}{#1}}
\newcommand{\DDDot}[1]{\stackrel{\mbox{...}}{#1}}
\newcommand{\DDDDot}[1]{\stackrel{\mbox{....}}{#1}}
\newcommand{\DDDDDot}[1]{\stackrel{\mbox{.....}}{#1}}

\begin{document}

\title[Timing models for the long-orbital period binary pulsar PSR B1259$-$63]
      {Timing models for the long-orbital period binary pulsar PSR B1259$-$63}

\author[N. Wex et al.]{N.~Wex$^{1,2,3}$, S.~Johnston$^2$, 
R.~N.~Manchester$^4$, A.~G.~Lyne$^5$, 
\newauthor B.~W.~Stappers$^6$ \& M.~Bailes$^7$\\
$^1$Max-Planck Society, Research Unit ``Theory of Gravitation'', 
Max-Wien-Pl.\ 1, D-07743 Jena, Germany \\
$^2$Research Centre for Theoretical Astrophysics, University of Sydney, 
NSW 2006, Australia \\
$^3$Max-Planck-Institut f\"ur Radioastronomie, Auf dem H\"ugel 69, 
D-53121 Bonn, Germany \\
$^4$Australia Telescope National Facility, CSIRO, PO Box 76, Epping, 
NSW 2121, Australia \\
$^5$University of Manchester, NRAL, Jodrell Bank, Macclesfield, 
Cheshire SK11 9DL, UK \\
$^6$Mount Stromolo and Siding Spring Observatories, ANU, Private Bag, 
Weston Creek ACT 2611, Australia \\
$^7$University of Melbourne, School of Physics, Parkville, 
Victoria 3052, Australia}

\maketitle

\begin{abstract} 
The pulsar PSR B1259$-$63 is in a highly eccentric 3.4-yr orbit with the Be
star SS 2883. Timing observations of this pulsar, made over a 7-yr period
using the Parkes 64-m radio telescope, cover two periastron passages, in 1990
August and 1994 January. The timing data cannot be fitted by the normal pulsar
and Keplerian binary parameters. 
A timing solution including a (non-precessing) Keplerian orbit and timing
noise (represented as a polynomial of fifth order in time) provide a
satisfactory fit to the data. However, because the Be star probably has a
significant quadrupole moment, we prefer to interpret the data by a
combination of timing noise, dominated by a cubic phase term, and $\dot\omega$
and $\dot x$ terms. We show that the $\dot\omega$ and $\dot x$ are likely to
be a result of a precessing orbit caused by the quadrupole moment of the
tilted companion star.  
We further rule out a number of possible physical effects which could
contribute to the timing data of PSR B1259$-$63 on a measurable level.
\end{abstract}

\begin{keywords} 

pulsar timing -- timing noise -- binary pulsars -- classical spin-orbit 
coupling -- pulsars: individual (PSR B1259$-$63) 

\end{keywords}


\section{Introduction}

The pulsar PSR B1259$-$63 is a member of a unique binary system. Discovered
using the Parkes telescope in a survey of the Galactic plane at 1.5 GHz
\cite{jlm+92a}, it was shown by Johnston et al.~(1992b) to be in a highly
eccentric 3.4-yr orbit with a 10th-magnitude Be star, SS 2883. The pulsar
period, $P$, is relatively short, 47.8 ms, and the measured period derivative,
$\dot{P}$, gives a pulsar characteristic age, $\tau_c = P/(2\dot{P})$, of $3.3
\times 10^5$ yr and a surface magnetic field of 3.3$\times 10^{11}$ G.  This
therefore is a young system, which may evolve through an accretion phase to
form a single or binary millisecond pulsar.  Rapidly spinning neutron stars
can only accrete matter if the co-rotation velocity at the Alfv\'en radius is
less than the Keplerian velocity at the same radius \cite{bv91}. Equality of
these velocities defines the `spin-up line'. At present, PSR B1259$-$63 lies
well to the left of the spin-up line, so that accretion onto the neutron star
is not possible until either the pulsar slows down or the pulsar magnetic
field decays.

Timing observations of PSR B1259$-$63, made over a 3.4-yr interval and
covering the 1990 August periastron, were reported by Johnston et al.\
\shortcite{jml+94}. A phase-connected fit to these data gave parameters for
the pulsar and its orbit, and showed that the next periastron would occur on
1994 January 9. This paper also reported optical observations which indicate
that the companion star is of spectral type B2e, with a mass of $m_* \sim 10$
M$_{\sun}$ and radius $R_* \sim 6$ R$_{\sun}$. The break-up velocity at the
equator, $v_{{\rm max}}$, for B2e stars is not very well known; it is
estimated to be $\sim 380$ km s$^{-1}$ by Slettebak et al.\ (1980) and $\sim
480$ km s$^{-1}$ by Schmidt-Kaler (1982).  Recent work by Porter (1996)
suggests that most Be stars rotate at $\sim 70$ per cent of the break-up
velocity.  From the mass function, a companion mass of 10 M$_{\sun}$ and a
pulsar mass, $m_p$, of 1.4 M$_{\sun}$ imply an orbital inclination
$i=36\degr$. The orbital eccentricity is very high, 0.87, and the pulsar
approaches within 24 $R_*$ of the companion star at periastron, passing
through the circumstellar disk. Extensive observations of the pulsar were made
at several radio frequencies before and after the 1994 January periastron, in
order to probe the circumstellar environment of SS 2883 (Johnston et al.\
1996; Melatos, Johnston \& Melrose 1995).

Observations made between 1990 January and 1994 October were well explained
by step changes in the pulsar period at the two periastrons \cite{mjl+95},
attributed to a propeller-torque spin-down caused by the
interaction of the pulsar with the circumstellar matter at the Alfv{\'e}n
radius (Illarionov \& Sunyaev 1975, King \& Cominsky 1994, Ghosh 1995).

In this paper we report on additional timing observations made using the
Parkes radio telescope over the past two years which, together with the
earlier data, give a total timing data span of seven years. We find that the
timing solution of Manchester et al. (1995) does not fit the recent data and
discuss alternative models and their interpretation.


\section{Timing observations and data analysis}

A total of $\sim$300 pulse times of arrival (TOAs) 
were measured at the Parkes radio
telescope between 1990 January and 1996 December. Most of the observations
were at frequencies around 1.5 GHz, but observations at 0.43, 0.66, 4.8 and
8.4 GHz were also made. At all frequencies, dual-channel cryogenically cooled
systems receiving orthogonal linear polarizations were used. After conversion
to an intermediate frequency, signals for each polarization were split into
sub-bands using filterbanks with channel widths of 0.125 or 0.25 MHz for
frequencies below 1 GHz, and 5 MHz for higher frequencies. Signals from
corresponding filter polarization pairs were detected, summed, high-pass
filtered and one-bit digitized, usually with a sampling interval of 0.6
ms. Further details of the observing systems are given in Johnston et al.\
\shortcite{jml+96}.

The data were folded at the topocentric period to form mean pulse profiles.
Integration times were typically 10 min at 1.5 GHz, and 20 to 30 min at other
frequencies.  Data taken prior to 1994 were convolved with a `standard
profile' in the time domain to produce TOAs whereas data taken after 1994 were
processed using a different method which involves matching the Fourier
components of the observed and standard profiles.  Uncertainties in the TOAs
are somewhat smaller using the second method.  However in order not to bias
the fitting procedure towards the second half of the data (i.e. the second
orbit), we use {\it unweighted} fits throughout this paper; that is, all TOAs
were assigned the same weight regardless of the estimated TOA uncertainty.
Also, we use only TOAs obtained at frequencies around 1.5 GHz and those TOAs
at 4.8 GHz, which show the best signal-to-noise ratio, in order to obtain a
good value for the dispersion measure. Pulsar and binary parameters were
obtained using the least-squares fitting program TEMPO \cite{tw89} with the
Jet Propulsion Laboratory solar-system ephemeris DE200 \cite{sta82}. TEMPO was
extended by the addition of a new timing model for binary pulsars that orbit a
companion star with a significant quadrupole moment (Wex 1998).

As discussed by Johnston et al.\ \shortcite{jml+96}, significant dispersion
and scattering changes were observed around periastron. Because of this, data
from 1990 July and 1993 December were omitted from the analysis.  The pulsar
was eclipsed during 1994 January. Thus, the two largest gaps in the data in
terms of true anomaly, $\varphi$, are one of 129 days around the first
periastron ($\varphi=-145\degr\dots100\degr$) and 76 days around the second
periastron ($\varphi=-128\degr\dots105\degr$). This will emerge to be a
problem when searching for the correct timing model for PSR B1259$-$63.

The data were first fitted for pulsar position, period, period derivative,
dispersion measure, and the five Keplerian orbital parameters. The best
results give an rms residual of 2100 $\mu$s and the post-fit residuals shown
in Fig.~1. Systematic variations in the residuals are observed at all orbital
phases, showing that this set of parameters does not model the
timing behaviour of the system satisfactorily.

\begin{figure}
\psfig{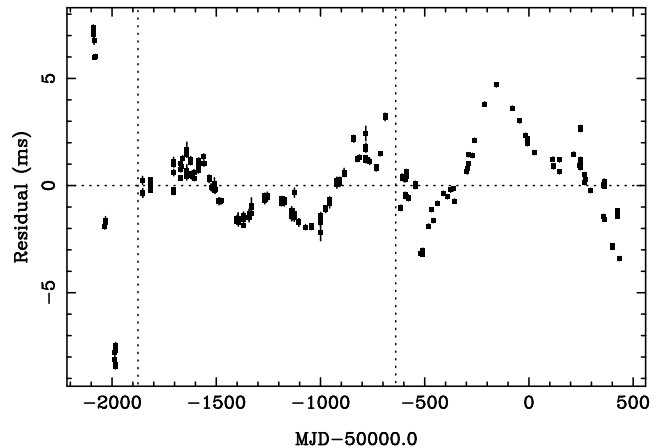} 
\caption{Post-fit residuals from an unweighted fit for pulsar position,
  period, period derivative, dispersion measure and Keplerian orbital
  parameters. The rms residual is 2100 $\mu$s. The vertical dotted lines in
  this and subsequent figures indicate the two epochs of periastron passage.}
\end{figure}

Three types of model are considered in the following subsections with a view to
minimising these residuals and understand their origin.

\subsection{Period steps at periastron}

As mentioned in the Introduction, observations made between 1990 January and
1994 October were well fitted by step increases in the pulsar period at the
two periastrons.  Fig.~2 shows the pre-fit residuals obtained using the
timing solution given in Manchester et al.\ \shortcite{mjl+95} on the
extended data set.  Clearly, the timing observations made in 1995 and 1996
are not modelled by this solution.

\begin{figure}
\psfig{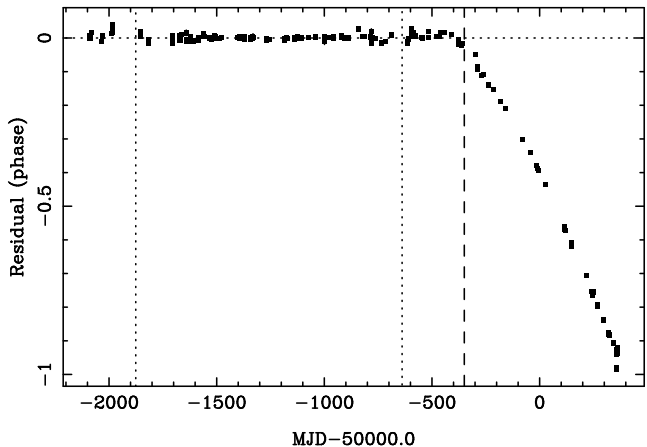} 
\caption{Pre-fit residuals for the Manchester et al.\ (1995) timing
         solution. The vertical dashed line indicates the end of the
         Manchester et al.\ (1995) data set.}
\end{figure}

There is other evidence which suggests that propeller spin-down does not occur
at periastron. Models of the system by Tavani \& Arons (1997) based on
extensive X-ray and $\gamma$-ray observations during the 1994 periastron
(Kaspi et al.\ 1995, Hirayama et al.\ 1996, Grove et al.\ 1995) conclude that
accretion or propeller-powered processes are ruled out.

If we allow fitting for a period step at the two periastrons, we find a rather
good fit, with rms residual of 340$\mu$s. However, the two period steps have
opposite sign, $\Delta\nu=-1.7\times10^{-8}$ at the first periastron and
$\Delta\nu=2.6\times10^{-8}$ at the second periastron. This does not seem
physically plausible and this evidence, coupled with the X-ray studies, causes
us to rule out this physical mechanism as an explanation for the timing
residuals in PSR B1259--63.

\subsection{Timing noise}

PSR B1259$-$63 is a comparatively young pulsar with a period derivative, $\Dot
P$, of $2.3\times10^{-15}$. Such a young pulsar should suffer timing noise
which is usually dominated by a cubic term and may contain higher-order
derivatives \cite{lyn96}. For a pulsar with such a period first derivative the
expected value of $\DDot P$ is of order $\pm2\times 10^{-26}$ s$^{-1}$. (The
intrinsic $\DDot P$ arising from magnetic dipole spin-down, i.e. assuming a
braking index of 3, is expected to be only of order $10^{-28}$ s$^{-1}$.)
Fitting for $\DDot P$, $\DDDot P$ and $\DDDDot P$ terms gives the residuals
illustrated in Fig.~3 and an rms residual of only 350 $\mu$s.  Parameters for
this fit are given in Table~1 as Model 1. The value for $\DDot P$ is within
the expected range. Fitting for a $\DDDDDot P$ does not improve the fit and
does not give significant values for this parameter.

The parameters in this fit are uncertain in the sense that they are relatively
insensitive to adding or subtracting an integral number of phase turns at the
two periastrons, provided that the number of turns added or subtracted was the
same at each periastron. This ambiguity arises because of the large ($\sim
100$ days) gap in the data resulting from the pulsar eclipse at each
periastron. Quoted parameter-error estimates take this uncertainty into
account by reflecting the parameter range for added or subtracted turns giving
satisfactory solutions, that is, increasing the rms residual by less than 20
per cent. Up to 3 phase turns could be added or subtracted at each periastron
within this criterion.

It is difficult to derive an `activity parameter' (Cordes \& Helfand 1980) for
binary pulsars, especially for long-period binaries where some of the timing
noise may be taken up in the orbital parameters. Having removed the lower
order terms in the fitting process above, the remaining residuals in Fig.\ 3
are probably the higher order terms of the noise process. In particular, the
timing residuals are very similar to those of PSR~B1951+32 \cite{flsb94},
a solitary pulsar of similar period and youth to PSR~B1259$-$63.

\begin{figure}
\psfig{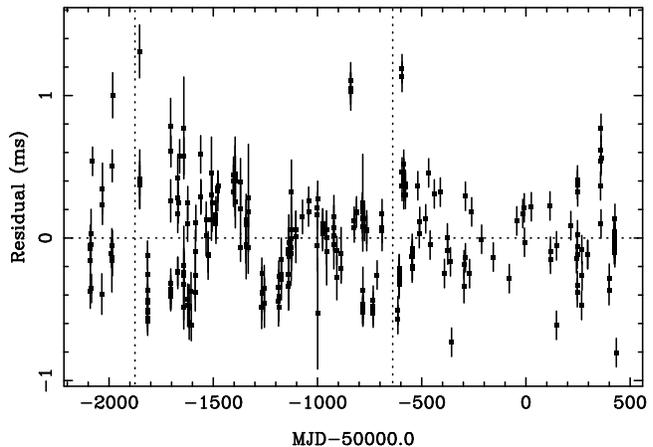}
\caption{Post-fit residuals for Model 1, i.e.\ fitting for pulsar 
   position, period, period derivative, dispersion measure, Keplerian orbital 
   parameters, and $\DDot P$, $\DDDot P$ and $\DDDDot P$ to model the timing 
   noise. The rms residual is 350 $\mu$s.}
\end{figure}

\subsection{Secular changes in the orbit ($\dot\omega$, $\dot x$) and
            classical spin-orbit coupling}

For a pulsar in orbit with a Be star, one may expect changes in the longitude
of the periastron, $\omega$, and the inclination of the orbit, $i$,
\cite{lbk+95}.  The latter manifests itself as a change of the projected
semi-major axis $x=a_p\sin i$. The physical cause of these changes is the
`classical spin-orbit coupling', i.e. the fact that the spin-induced
quadrupole of the fast-rotating companion leads to a $1/r^3$ term in the
gravitational potential which in turn leads to apsidal motion and precession
of the binary orbit (Kopal 1978, Smarr \& Blandford 1976, Lai et al.\ 1995,
Wex 1998).

Two unweighted fits for the pulsar-spin parameters, including a $\ddot P$ to
model the long-term behaviour of the timing noise, the Keplerian parameters
for orbital motion, and the two post-Keplerian parameters, $\dot\omega$ and
$\dot x$, are given in Table~1 (Model 2A and Model 2B). The corresponding
post-fit residuals are shown in Fig.~4. Again, realistic errors for the
parameters in the table were obtained by adding (or subtracting) an integer
number of phase turns at each periastron until the fit deviated from the best
fit by about 20 per cent. In this case, however, two satisfactory independent
fits could be obtained by having, in the case of Model 2A, an equal number of
phase turns at the two periastrons and, in the case of Model 2B, one more
phase turn at the second periastron.  

\begin{figure}
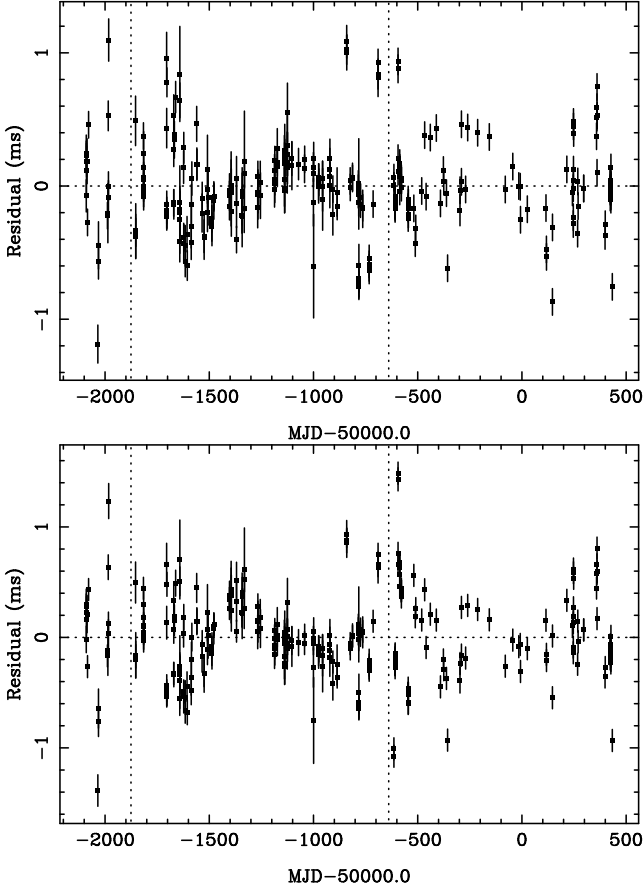

\psfig{figure=fig4a.ps,angle=-90,width=8.5cm}
\psfig{figure=fig4b.ps,angle=-90,width=8.5cm} 
\caption{Post-fit residuals for Model 2A (upper) and Model 2B (lower), i.e.\
         fitting for pulsar position, period, period derivative, dispersion
         measure, Keplerian orbital parameters, $\Dot\omega$, $\Dot x$ and
         $\DDot P$ to model the timing noise, using the BT++ timing model (see
         Wex 1998). The two models differ in the number of turns added at
         periaston (see text). Model 2A has an rms residual of 340 $\mu$s and
         Model 2B an rms residual of 390$\mu$s.}
\end{figure}

Let us assume that all the timing noise is modelled by $\DDot P$, i.e. that
the fitted $\Dot\omega$ and $\Dot x$ result only from changes in the
orientation of the binary orbit due to classical spin-orbit coupling, and
focus on the question: can the values for $\Dot\omega$ and $\Dot x$, as
given in Table~1, be explained by a reasonable quadrupole moment and
orientation of the Be star?

\begin{figure}
\psfig{figure=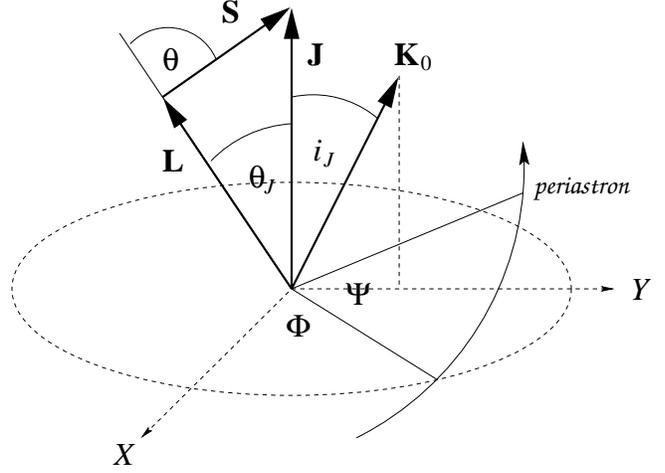,width=8.5cm} 
\caption{Definition of different angles in the binary system. The invariable
         plane ($X$-$Y$) is perpendicular to the total angular momentum
         $\mbf{J} = \mbf{L} + \mbf{S}$, and the line-of-sight unit vector
         $\mbf{K}_0$ lies in the $Y$-$Z$ plane.  $\theta_J$ is the inclination
         of the orbital plane with respect to the invariable plane, $\Phi$ is
         the longitude of the ascending node with respect to the invariable
         plane and $\Psi$ is the longitude of the periastron with respect to
         the invariable plane. $\theta$ is the inclination of the companion
         spin with respect to the orbital angular momentum. $i_J$ and $i$ are
         the angles between the line of sight and $\mbf{J}$ and $\mbf{L}$
         respectively.}
\end{figure}

In the case of PSR B1259$-$63, the spin angular momentum of the pulsar is
completely negligible compared to the spin angular momentum $\mbf{S}$ of the
companion and the (total) orbital angular momentum $\mbf{L}$. The angle
$\theta_J$ between the conserved total angular momentum $\mbf{J} = \mbf{L} +
\mbf{S}$ and $\mbf{L}$ is also conserved (see Fig.~5).  The intersection of
the orbital plane (perpendicular to $\mbf{L}$) and the invariable plane
(perpendicular to $\mbf{J}$) has the longitude $\Phi$ and the longitude of the
periastron with respect to the invariable plane is denoted by $\Psi$. The
angle between $\mbf{L}$ and the direction of sight $\mbf{K}_0$ is denoted by
$i$. The angle $i$ is not uniquely determined by timing observations and is
equal to either 36\degr or 180\degr--36\degr for PSR B1259--63.

The classical spin-orbit coupling causes a (linear-in-time) precession of the
angles $\Phi$ and $\Psi$ which implies a rather complicated change of the
angles $i$ and $\omega$.  For the case $|\mbf{S}|\ll|\mbf{L}|$
\footnote[2]{For PSR B1259$-$63, $|\mbf{S}|\sim 0.1|\mbf{L}|$. The
approximation $|\mbf{S}|\ll|\mbf{L}|$ is justified for the accuracy required
here.} the changes of the longitude of periastron, $\omega$, and the projected
semi-major axis, $x$, are approximated by
\be\ba{l}\label{omdot}
  \DS \dot\omega_Q \simeq \frac{3n^{7/3}\tilde J_2^*}{2(GM)^{2/3}(1-e^2)^2}
      \times \\[4mm] \DS \qquad\qquad \times
      \left(1-\frac{3}{2}\sin^2\theta+\cot i\sin\theta\cos\theta\cos\Phi
      \right) \,, 
\ea\ee\be\label{xdot}
  \frac{\dot x_Q}{x} \simeq \frac{3n^{7/3}\tilde J_2^*}{2(GM)^{2/3}
      (1-e^2)^2}\,\cot i \sin \theta \cos \theta \sin\Phi \,, 
\ee
(Smarr \& Blandford 1976, Lai et al.\ 1995, Wex 1998). $n\equiv 2\pi/P_b$ is
the orbital frequency of the binary system, $G$ is the gravitational constant,
and $M=m_p+m_*$ the total mass of the system.  The quantity $\tilde J_2^*$ is
a measure of the quadrupole moment of the Be star
\be\label{J2}
   \tilde J_2^*\equiv\frac{I_3^*-I_1^*}{m_*}=\frac{2}{3}kR_*^2
   \hat\Omega_*^2 \,,
\ee
where $\hat\Omega_*\equiv\Omega_*/(Gm_*/R_*^3)^{1/2}$ is equal to unity if the
star is rotating near break-up. $I_3^*$ and $I_1^*$ are the moments of inertia
about the spin axis and an orthogonal axis, respectively. $R_*$ is the
(equatorial) radius of the Be star and $k$ is the apsidal motion constant
which is of the order of 0.01 for a 10 $M_\odot$ main-sequence star
(Schwarzschild 1958; Claret \& Gimenez 1992). Assuming a typical radius of
$R_*\approx 6R_\odot$ and a rotation of $\hat\Omega_*\approx0.7$ for the Be-star 
companion we find
\be\label{Q}
   \tilde J_2^* \sim 0.1 R_\odot^2 = 2.2\times10^{-6}\:{\rm AU}^2\,.
\ee
The ratio of equations (\ref{omdot}) and (\ref{xdot}) leads to 
\be\label{ratio}
   \frac{\dot\omega_Q x}{\dot x_Q}\sin\Phi-\cos\Phi = 
   \frac{1+3\cos2\theta}{2\cot i\sin 2\theta} \,,
\ee
an equation that does not contain the uncertain quantity $\tilde J_2^*$.

The observed value of $\dot\omega$ (see Table~1) is a compound of
$\dot\omega_Q$ and the general relativistic contribution
\be
   \dot\omega_{GR} \approx +0.00003\:{\rm deg}\:{\rm yr}^{-1} \,.
\ee
Thus we have for PSR B1259$-$63
\be 
\mbox{\sc Model 2A}: \quad \left\{\ba{l}
   \dot\omega_Q = -0.00021(1)\:{\rm deg}\:{\rm yr}^{-1} \\[2mm]
   \dot x_Q     = -0.15(3) \times 10^{-10}
\ea\right.
\ee
\be 
\mbox{\sc Model 2B}: \quad \left\{\ba{l}
   \dot\omega_Q = -0.00043(1)\:{\rm deg}\:{\rm yr}^{-1} \\[2mm]
   \dot x_Q     = -2.40(4) \times 10^{-10}
\ea\right.
\ee
The restrictions implied by equation~(\ref{ratio}) are plotted in Fig.~6
which, in addition, shows restrictions implied by optical observations
of the projected stellar rotation velocity, $v_{{\rm obs}}$:
\be\label{optic}
   \pm\sqrt{1-\left(\frac{ v_{{\rm obs}} }{ \hat\Omega_* v_{{\rm max}} }
   \right)^2}=\cos i \cos \theta+ \sin i \sin \theta\cos\Phi, 
\ee
where $v_{{\rm max}}$ is the break-up velocity. For SS 2883 one finds 180 km
s$^{-1}$ for $v_{{\rm obs}}$ (Johnston et al.\ 1994). As mentioned in Section
1, the break-up velocity for Be stars is not very well known; we use $v_{{\rm
max}}=400\:{\rm km/s}$ as the break-up velocity of SS 2883.  If we assume that
SS 2883 rotates at 70\% of its break-up velocity ($\hat\Omega_*=0.7$) we get
the dotted line in Fig.\ 6 and allowed values of ($\theta$, $\Phi$) are
therefore close to (75\degr, 355\degr) and (105\degr,175\degr).

\begin{figure}
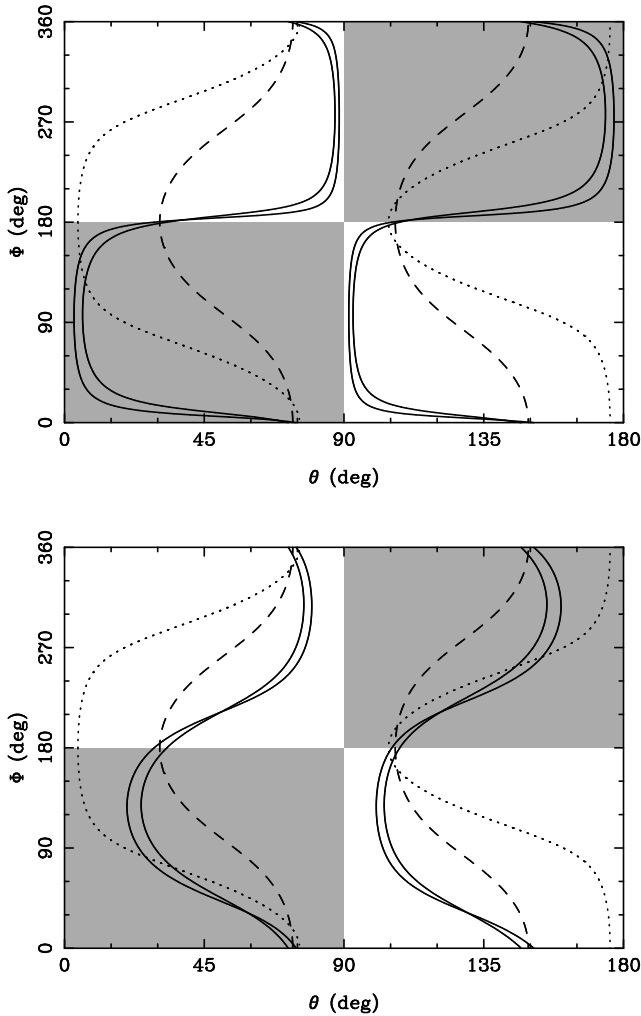

\psfig{figure=fig6a.ps,angle=-90,width=8.5cm}
\vspace{5mm}
\psfig{figure=fig6b.ps,angle=-90,width=8.5cm}
\caption{$\theta$-$\Phi$ parameter space for Model 2A (upper) and Model 2B
         (lower) for the case $i=36\degr\pm5\degr$. (for $i=144\degr\pm5\degr$
         one has to replace $\Phi$ by $\Phi-180\degr$). The grey areas are
         excluded because $\dot x<0$. The regions to the left of the left
         dashed curve and to the right of the right dashed curve are excluded
         because $\dot\omega_0<0$. The solid curves border the narrow region
         of possible values of $\theta$ and $\Phi$. For the regions to the
         left of the left dotted curve and to the right of the right dotted
         curve the Be star's rotation exceeds 70\% of the break-up velocity.}
\end{figure}

The large value of $\theta$ implies a significant birth kick to the neutron
star (cf.\ Kaspi et al.\ 1996). Since recent measurements of pulsar velocities
suggest kicks of 200 -- 400 km s$^{-1}$ are commonplace (Lyne \& Lorimer
1994), there is a high probability that PSR B1259--63 was kicked into a highly
inclined orbit; indeed, there is a high probability that an eccentric orbit
resulting from a birth kick is retrograde (Hills 1983).

Using equations (\ref{omdot}) and (\ref{ratio}) one can calculate $\tilde
J_2^*$ as a function of $\Phi$, and so using equation (\ref{J2}) one obtains
$kR_*^2$ as a function of $\Phi$.  Assuming that $R_*\approx 6R_\odot$ for B2e
stars, we find those values for the apsidal motion constant $k$ which are
necessary to explain the parameters $\dot\omega$ and $\dot x$ for Models 2A
and 2B by classical spin-orbit coupling exclusively. We find that Models 2A
and 2B need rather large values for the apsidal motion constant $k$. While the
minimum value for Model 2A, $k=0.02$, is still in the range of theoretical
models (Claret \& Gim{\'e}nez 1991), the value of $k=0.05$, necessary for
Model 2B, seems to be too large. 

As in the previous section, this model provides a good explanation of the
data, i.e.\ this solution maintaines phase to better than 3\% of the
rotational period.  The remaining residuals are fully consistent with higher
order terms of the timing noise (c.f.\ end of Section 2.2). However, it is
likely in this case that such timing noise cannot fully be removed with a
$\DDot P$ term and that this may affect the measured values of $\dot\omega$
and $\dot x$ and hence the derived values for $\theta$ and $\Phi$.  This
introduces an additional uncertainty which is not represented by the quoted
error limits.

Simulations suggest that observations around and after the next periastron
(1997 May 29) should distinguish between the various models. 


\section{Further effects on the orbital motion}

Besides quadrupolar effects resulting from the oblateness of the Be star,
orbital perturbations could arise from tidal effects on the Be star,
frictional drag as the pulsar passes through the circumstellar disk, and mass
loss from the Be star. We will show that the major influence of these effects
is to change the orbital period, $P_b$, by at most 3 seconds per orbital
period ($\dot P_b\la 3\times10^{-8}$). However if we additionally fit for
$\dot P_b$ in the models described above, the rms does not change (i.e.\ the
value of $\dot P_b$ is not significant) and the 1$\sigma$ errors on $\dot P_b$
are of order $10^{-5}$. Hence none of these effects have any impact on the
current timing solution at a measureable level.

\subsection{Tidal effects}

In the case of PSR B1259$-$63, perturbations due to tidal effects were
investigated by several authors using standard theory of tidal dissipation and
found to be negligible (Kochanek 1993, Manchester et al.\ 1995, Lai et
al.\ 1995).

Resent work by Lai (1996,1997) and Kumar \& Quataert (1997) shows that
(differential) stellar rotation can change the strength of the dynamical tide
significantly. In particular, retrograde rotation (with respect to the orbital
motion of the pulsar) increases the energy and angular-momentum transfer by
more than two orders of magnitude. This effect is most likely the explanation
for the observed $\dot P_b$ of $3\times10^{-7}$ in the main-sequence star
binary pulsar PSR J0045--7319 (Kaspi et al.\ 1996).  However the periastron
distance of PSR J0045--7319 from its companion is only 4 stellar radii, while
PSR B1259--63 passes periastron at a distance of 24 stellar radii and thus the
strength of the dynamical tides is several orders of magnitude smaller.
Following calculations in Lai (1996) we find for PSR B1259--63
\be\label{dyntide}
   \dot P_b \sim -10^{-10} \left(\frac{50\:{\rm yr}}{\tau_d}\right)
   \left(\frac{T_2(\eta)}{0.001}\right) \,,
\ee
where $\tau_d$ is the dissipation time for the dynamical tide and $\eta =
(m_*/M)^{1/2}(r_{{\rm peri}}/R_*)^{3/2}$ where $r_{{\rm peri}}$ is the
periastron distance. The function $T_2(\eta)$ depends strongly on the rotation
rate of the companion star and increases with decreasing values for
$\hat\Omega_*$.  Assuming nearly maximum retrograde rotation of the companion
star, Lai (1997) found $T_2=0.01$ for PSR J0045--7319 ($\eta=7$). For PSR
B1259--63, $T_2$ should be clearly smaller, since $\eta=110$ for this binary
system; one expects $T_2\la0.001$. Therefore, for realistic values of the
dissipation time, $\tau_d$, the dynamical tides are absolutely negligible in
the PSR B1259--63 system.

\subsection{Frictional drag}

A fit to dispersion-measure changes around the 1994 January periastron 
\cite{jml+96} indicates that the radial dependence of the electron density
in the circumstellar disk of SS 2883 has the form
\be
   n_e(r) \sim 4.5\times10^{12}(r/R_*)^{-4.2}\:{\rm cm}^{-3}\,.
\ee 
Assuming a hydrogen plasma, this corresponds to a mass density
\be\label{rho}
   \rho_H(r) \sim 7.5\times10^{-12}(r/R_*)^{-4.2}\:{\rm g}\:{\rm cm}^{-3}\,.
\ee
The radius within which matter is captured by the pulsar is given by the 
Bondi \shortcite{bon52} relation
\be
   r_{\rm acc} = \frac{2Gm_p}{v_{\rm rel}^2} \approx \frac{3.7 \times 
                   10^{12}\,{\rm cm}}{(v_{\rm rel}/100\,{\rm km\;s}^{-1})^2}\,,
\ee
where $v_{\rm rel}$ is the velocity relative to the surrounding medium.
The drag acceleration induced by the accretion of matter onto the neutron star
is  then given by
\be
   \mbf{A} = -\frac{\pi r_{{\rm acc}}^2\varrho_H v_{\rm rel}\mbf{v}_{\rm rel}}
              {m_p} \,.
\ee
To estimate if friction has an important effect on the orbital motion of PSR
B1259$-$63, we assume that the circumstellar disk is not inclined with respect
to the orbital plane, implying a maximum effect. We further assume (to
simplify the calculations) that the velocity of the pulsar around periastron
($\sim 150$ km s$^{-1}$) is much larger than the velocity of the circumstellar
material\footnote[3]{The velocity of the circumstellar matter is expected to
be of the order of 150 to 300 km s$^{-1}$ (Waters 1986; Bjorkman \& Cassinelli
1993). Although we neglect this fact, we will still obtain the correct order
of magnitude.}. In this case the direction of the friction is roughly
anti-parallel to the velocity of the pulsar, $\mbf{v}_{{\rm rel}} =
\mbf{v}_{{\rm p}}$. Friction will not affect the plane of the orbit, but it
will affect the Keplerian parameters. For the change of the orbital period
$P_b$, eccentricity $e$, and longitude of periastron $\omega$ one finds (see
e.g.\ Danby 1962)
\be
   \dot P_b    = \frac{3P_b^{4/3}v}{(2\pi GM)^{2/3}}\,A \,,
\ee\be
   \dot \omega = \frac{2\sin\varphi}{ev}\,A \,,
\ee\be
   \dot e      = 2\,\frac{\cos\varphi+e}{v}\,A \,,
\ee
where $\varphi$ is the true anomaly of the pulsar and $v$ the absolute value
of the relative velocity between the pulsar and its companion. Integration of
these equations leads to the following change of the orbital parameters during
one orbital revolution: 
\be\label{pbf}
   \Delta P_b \approx -0.002 \: {\rm s} \,,
\ee\be
   \Delta \omega = 0 \,,  
\ee\be\label{ef}
   \Delta e \approx -3\times10^{-13} \,.
\ee
These values are well below the present measurement precision. Thus we
conclude, in agreement with Manchester et al.\ (1995), that frictional drag
has a negligible effect on the TOAs of PSR B1259$-$63.

>From equation~(\ref{ef}), we can calculate the typical time scale for
circularisation:
\be 
   \tau_{{\rm circ}} \ga \frac{e}{\Delta e}P_b \sim 10^{13}\:{\rm yr} \,.
\ee
There are indications that the circumstellar disk is tilted with respect to
the orbital plane \cite{jml+96}. Therefore the pulsar is within the
circumstellar disk twice per orbit, but only for a comparatively short time,
which reduces the effect of friction by at least an order of magnitude.
Furthermore there is no evidence that any accretion of material onto the
pulsar took place during the 1994 periastron passage (Tavani \& Arons 1997).

\subsection{Mass loss of the companion}

If the companion star loses mass at a rate $\dot m_*$ so that there is no 
linear momentum loss in the instantaneous rest frame of the star, the orbital
period will change by a rate given by (Jeans 1924, 1925)
\be
    \frac{\dot P_b}{P_b} = -\frac{2 \dot m_*}{m_* + m_p}\,. 
\ee
The mass loss of SS 2883 is of order $-5\times10^{-8} M_\odot$ yr$^{-1}$
\cite{jml+96} and so
\be
    \dot P_b \sim 3 \times 10^{-8} \,, 
\ee
which gives an orbital period change of $\sim 3$ seconds per orbital
revolution, a value which is well below measurement precision.


\section{Relativistic effects}

Because of the high eccentricity and the large mass of the companion there
are two relativistic effects which should be considered: the Einstein delay
described by the post-Keplerian parameter $\gamma$ \cite{bt76}, and the
Shapiro delay, which depends on the orbital inclination $i$ of the binary
system and is proportional to the mass of the companion \cite{dd86}.

For PSR B1259$-$63, the parameter $\gamma$ is quite large. We find 
\be 
   \gamma = \frac{1}{c^2} \left(\frac{P_b}{2\pi}\right)^{1/3} \frac{G^{2/3}
   m_* (m_p + 2m_*)}{(m_p + m_*)^{4/3}}\; e \simeq 0.54 \: {\rm s} \,.  
\ee
As pointed out by Blandford \& Teukolsky (1975) and Brumberg et al. (1975),
$\gamma$ can be isolated only in the presence of apsidal motion and requires
observations over a time interval in which $\omega$ changes by a significant
amount. Since, for PSR B1259$-$63, $\omega$ changes by a few arcseconds each
orbital revolution and less than two orbital revolutions have been observed,
we cannot extract $\gamma$ from the present data; it is absorbed into the
Keplerian parameters $x$ and $\omega$. To a first approximation we find
\bea 
   x &\longrightarrow& x+\frac{\gamma\cos\omega}{(1-e^2)^{1/2}}
   \simeq x - 0.84 \: {\rm sec} \,, \\ \omega &\longrightarrow&
   \omega-\frac{\gamma\sin\omega}{x(1-e^2)^{1/2}} \simeq \omega - 0\fdg031\,.
\eea
Thus the observed values are offset from the true values by a few hundred
times the errors of the measurements.

A second potentially important relativistic effect is the Shapiro delay, the
propagation delay caused by the gravitational potential of the companion
star. The delay is given by \cite{dd86}
\be\ba{l}
   \Delta_S = \DS -\frac{2Gm_*}{c^3}\ln\left\{1-e\cos U-[\sin\omega(\cos U-e)
              \right. \\ \qquad\qquad\qquad\qquad\quad\:
              \left.+(1-e^2)^{1/2}\cos\omega\sin U]\sin i\right\}\,,
\ea\ee
where $U$ is the eccentric anomaly of the binary orbit. For PSR B1259$-$63
($i\approx 35\degr$) $\Delta_S$ has a range of 400 $\mu$s and thus is much
greater than the error in most of the TOAs. The fact that there are no
observations around periastron where there is a sharp peak in $\Delta_S$
reduces the actual span of $\Delta_S$ to 150 $\mu$s. This remaining effect
can be absorbed in the other parameters without changing them by more than
the typical given error (see Table~1). 


\section{Conclusions}

At present, the timing observations for the binary pulsar PSR B1259$-$63 span
seven years. Because of the gaps in timing observations around the two
periastrons and the large timing noise present in this young pulsar, we still
are not able to derive a unique timing model to explain the TOAs.  

Model 1 is a timing solution including a non-precessing Keplerian orbit and
timing noise represented as a polynomial of fifth order in time. This model
provides a satisfactory fit to the data. The remaining timing residuals are
understood as short-term timing noise similiar to that seen in
observations of other young pulsars (cf.\ Foster et al.\ 1994).

Equally good results were obtained by Model 2 and 3. Both timing models
contain just a $\ddot P$ term to account for the long-term behaviour of the
timing noise, and $\dot\omega$ and $\dot x$, which both are understood to
result from a precession of the orbit. This orbital precession can be
explained by the classical spin-orbit coupling caused by the quadrupolar
nature of the main-sequence star companion. The corresponding advance of
periastron is negative and thus the companion should be tilted by more than
$30\degr$ with respect to the orbital plane (See Fig.\ 6). This can be
explained by a birth kick for the pulsar (cf.\ Kaspi et al.\ 1996).

Tidal dissipation and frictional drag in the circumstellar matter is shown to
be negligible. The influence of the mass loss of the companion is too small to
be detectable unless the mass loss is $\ga 10^{-5} M_\odot$/yr. At the same
time we can exclude a significant orbital period change in the TOAs of PSR
B1259--63.

PSR B1259$-$63 should show the largest Einstein delay and largest Shapiro
delay of all known binary pulsars. The small change in the longitude of
periastron makes it impossible to isolate the Einstein delay. The Shapiro
delay peaks sharply around periastron and is so far unobservable.

Again, we stress that the physical parameters given here for the companion
star, $\theta$, $\Phi$ and $k$, should be understood as one possible
explanation for the significant values of $\dot\omega$ and $\dot x$ in Models
2A and 2B. If the long-term behaviour of the timing noise of PSR B1259$-$63 is
not fully modelled by a cubic term ($\ddot P$), it is possible that rather
large fractions of these parameters are not explained by a precession of the
orbital plane but have their origin in unmodelled timing noise. The parameters
here show that, in principle, all of the $\dot\omega$ and $\dot x$ can arise
from classical spin-orbit coupling, for Model 2A in particular.

Although Model 1 gives a good fit to the TOAs without making use of the
classical spin-orbit coupling, it seems unlikely that the classical
spin-orbit coupling is of no importance for this system.


\section*{Acknowledgments}

We thank the Parkes staff for their assistance, N. D'Amico, B. Koribalski
and L. Nicastro for help with these observations and J. H. Taylor for useful
comments. The Australia Telescope is funded by the Commonwealth of Australia
for operation as a National Facility managed by the CSIRO.




\clearpage
\newpage
\begin{table*}
\begin{minipage}{12.2cm}
\caption{Parameters for the three timing solutions for PSR B1259$-$63, Model 1
         ($\DDDot P$, $\DDDDot P$), Model 2A ($\dot\omega$, $\dot x$), and
         Model 2B ($\dot\omega$, $\dot x$). The data span is MJD 47909 --
         50448. To estimate the parameter errors we compared all solutions for
         various combinations of integral numbers of phase turns at the two
         periastrons which give an rms residual not worse than 1.2 times the
         quoted rms residual. The error quoted in the table is the maximum
         difference between the values given in the table and the values
         obtained while changing the phase turns. Compared to the errors given
         by TEMPO and the errors obtained by the bootstrap method, the errors
         given here, in most cases, are more conservative.}
\begin{tabular}{llll}\hline
& Model 1 & Model 2A & Model 2B \\ \hline
Rms residual ($\mu$s)       & 350
	                    & 340 
                            & 390 \\
\\
$\alpha$ (J2000)            & $13^{\rm h}02^{\rm m}47\fs65(1)$
			    & $13^{\rm h}02^{\rm m}47\fs66(1)$	 
			    & $13^{\rm h}02^{\rm m}47\fs65(1)$ \\	 
$\delta$ (J2000)            & $-$63\degr 50\arcmin 08\farcs7(1) 
		            & $-$63\degr 50\arcmin 08\farcs7(1) 
		            & $-$63\degr 50\arcmin 08\farcs7(1) \\
$P$ (ms)                    & 47.7620537(1)
                            & 47.7620537(1)
                            & 47.7620537(1) \\
$\Dot P$ ($10^{-15}$)       & 2.2785(2)
			    & 2.27709(1) 
			    & 2.27720(1) \\
Epoch (MJD)                 & 48053.440 
		            & 48053.440 
		            & 48053.440 \\
DM (cm$^{-3}$pc)            & 146.78(5)
			    & 146.81(5)
			    & 146.85(5) \\
\\
$P_b$ (d)                   & 1236.7238(1) 
			    & 1236.7231(1)
			    & 1236.7244(1) \\
$x$ (s)                     & 1296.4(3)
			    & 1296.3(2)
			    & 1296.4(3) \\
$\omega$ (deg)              & 138.668(7) 
			    & 138.667(6) 
			    & 138.670(7) \\
$e$                         & 0.86990(4) 
		            & 0.86989(3) 
		            & 0.86991(4) \\
$T_0$ (MJD)                 & 48124.348(4) 
                            & 48124.350(4) 
                            & 48124.348(4) \\
\\
$\DDot P$ ($10^{-26}$ s$^{-1}$)   & $-$5.2(8)
                                  & $-$0.884(9) 
                                  & $-$0.788(1) \\
$\DDDot P$  ($10^{-34}$ s$^{-2}$) & $+$7(2) & --- & --- \\ 
$\DDDDot P$ ($10^{-42}$ s$^{-3}$) & $-$5(2) & --- & --- \\
$\Dot\omega$ (deg yr$^{-1}$)      & --- & $-$0.000184(8) & $-$0.000396(9) \\ 
$\Dot x$ ($10^{-10}$)             & --- & $-$0.15(3)    & $-$2.40(4)    \\

Epoch of Third periastron (UT)    & 1997/5/29 $19^{{\rm h}}$  
                                  & 1997/5/29 $19^{{\rm h}}$ 
                                  & 1997/5/29 $19^{{\rm h}}$ \\ 
\hline
\end{tabular}
\end{minipage}
\end{table*}


\begin{thebibliography}{}
  \bibitem[\protect\citename{Barker \& O'Connell }1975]{boc75}
    Barker B.~M., O'Connell R.~F., 1975, Phys.\ Rev.\ D, 12, 329
  \bibitem[\protect\citename{Bhattacharya \& {van den Heuvel} }1991]{bv91}
    Bhattacharya D., van den Heuvel E. P.~J., 1991, Phys.\ Rep., 203, 1
  \bibitem[\protect\citename{Bjorkman \& Cassinelli }1993]{bc93}
    Bjorkman J.~E., Cassinelli J.~P., 1993, ApJ, 409, 429
  \bibitem[\protect\citename{Blandford \& Teukolsky }1975]{bt75}
    Blandford R., Teukolsky, S. A., 1975, ApJ, 198, L27
  \bibitem[\protect\citename{Blandford \& Teukolsky }1976]{bt76}
    Blandford R., Teukolsky, S. A., 1976, ApJ, 205, 580
  \bibitem[\protect\citename{Bondi }1952]{bon52}
    Bondi~H., 1952, MNRAS, 112, 195
  \bibitem[\protect\citename{Brumberg et al.\ }1975]{bzn+75}
    Brumberg V. A., Zel'dovich Y. B., Novikov I. D., Shakura N. I., 1975,
    Astr.\ Lett., 1, 5
  \bibitem[\protect\citename{Claret \& Gimenez }1992]{cg92}
    Claret A., Gimenez A., 1992, A\&AS, 96, 255
  \bibitem[\protect\citename{Cordes \& Helfand }1980]{ch80}
    Cordes~J.~M., Helfand~D.~J., 1980, ApJ, 239, 640
  \bibitem[\protect\citename{Damour \& Deruelle }1985]{dd85}
    Damour T., Deruelle N., 1985, Ann.\ Inst.\ Henri Poincar{\'e}, 43, 107
  \bibitem[\protect\citename{Damour \& Deruelle }1986]{dd86}
    Damour T., Deruelle N., 1986, Ann.\ Inst.\ Henri Poincar{\'e}, 44, 263
  \bibitem[\protect\citename{Damour \& Taylor }1992]{dt92}
    Damour T., Taylor J. H., 1992, Phys.\ Rev.\ D, 45, 1840
  \bibitem[\protect\citename{Danby }1962]{dan62}
    Danby J. M. A., 1962, Fundamentals of Celestial Mechanics, The MacMillan
    Company, New York
  \bibitem[\protect\citename{Foster et al.\ }1994]{flsb94}
    Foster~R.~S., Lyne~A.~G., Shemar~S.~L., Backer~D.~C., 1994, AJ, 108, 175
  \bibitem[\protect\citename{Ghosh }1995]{gho95}
    Ghosh~P., 1995, ApJ, 453, 411
  \bibitem[\protect\citename{Hill }1983]{hil83}
    Hills J. G., 1983, ApJ, 267, 322
  \bibitem[\protect\citename{Hirayama et al.\ }1996]{hnt+} 
    Hirayama~M., Nagase~F., Tavani~M., Kaspi~V.~M., Kawai~N., 
    Arons~J., 1996, PASJ, 48, 833
  \bibitem[\protect\citename{Illarionov \& Sunyaev }1975]{is75}
    Illarionov~A.~F., Sunyaev R.~A., 1975, A\&A, 39, 185
  \bibitem[\protect\citename{Jeans 1924 }1924]{jea24}
    Jeans J.~H., 1924, MNRAS, 84, 2
  \bibitem[\protect\citename{Jeans 1925 }1925]{jea25}
    Jeans J.~H., 1925, MNRAS, 85, 912
  \bibitem[\protect\citename{Johnston et al.\ }1992a]{jlm+92a}
    Johnston~S., Lyne~A.~G., Manchester~R.~N., Kniffen~D.~A., D'Amico~N.,
    Lim~J., Ashworth~M., 1992a, MNRAS, 255, 401
  \bibitem[\protect\citename{Johnston et al.\ }1992b]{jlm+92b}
    Johnston~S., Manchester~R.~N., Lyne~A.~G., Bailes~M., Kaspi~V.~M., 
    Qiao~G., D'Amico~N., 1992b, ApJ, 387, L37.
  \bibitem[\protect\citename{Johnston et al.\ }1994]{jml+94}
    Johnston~S., Manchester~R.~N., Lyne~A.~G., Nicastro~L., Spyromilio~J.,
    1994, MNRAS, 268, 430
  \bibitem[\protect\citename{Johnston et al.\ }1996]{jml+96}
    Johnston~S., Manchester~R.~N., Lyne~A.~G., D'Amico~N., Bailes~M.,
    Gaensler~B.~M., Nicastro L., 1996, MNRAS, 279, 1026
  \bibitem[\protect\citename{Kaspi et al.\ }1996]{kbm+96}
    Kaspi~V.~M., Bailes~M., Manchester~R.~N., Stappers~B.~W., Bell~J.~F.,
    1996, Nature, 381, 584
  \bibitem[\protect\citename{Kaspi et al.\ }1995]{ktn+}
    Kaspi~V.~M., Tavani~M., Nagase~F.~D., Hirayama~M., Hoshino~M., 
    Aoki~T., Kawai~N., Arons~J., 1995, ApJ, 452, 819
  \bibitem[\protect\citename{King \& Cominsky }1994]{kc94}
    King~A., Cominsky~L., 1994, ApJ, 435, 411
  \bibitem[\protect\citename{Kochanek }1994]{koc93}
    Kochanek~C.~S., ApJ, 406, 638
  \bibitem[\protect\citename{Kopal }1978]{k78}
    Kopal~Z., 1978, Dynamics of Close Binary Systems, 
    D.~Reidel Publishing Company, Dordrecht, Holland
  \bibitem[\protect\citename{Kumar \& Quataert }1997]{kq97}
    Kumar~P., Quataert~E.~J., 1997, ApJ, 479, L51
  \bibitem[\protect\citename{Lai }1996]{lai96}
    Lai~D., 1996, ApJ, 466, L35
  \bibitem[\protect\citename{Lai }1997]{lai97}
    Lai~D., 1998, ApJ, Submitted, astro-ph/9704132
  \bibitem[\protect\citename{Lai et al.\ }1995]{lbk+95}
    Lai~D., Bildsten~L., Kaspi~V.~M., 1995, ApJ, 452, 819
  \bibitem[\protect\citename{Lyne }1996]{lyn96}
    Lyne~A.~G., 1996, Pulsars: Problems \& Progress, ASP Conference
    Series, S.~Johnston, M.~A.~Walker and M.~Bailes, eds.
  \bibitem[\protect\citename{Lyne \& Lorimer }1996]{ll94}
    Lyne~A.~G., Lorimer~D.~R., 1994, Nature, 369, 127
  \bibitem[\protect\citename{Manchester et al.\ }1995]{mjl+95}
    Manchester~R.~N., Johnston S., Lyne A.~G., D'Amico N., Bailes M.,
    Nicastro~L., 1995, ApJ, 445, L137
  \bibitem[\protect\citename{Melatos, Johnston, \& Melrose }1995]{mjm95}
    Melatos~A., Johnston~S., Melrose D.~B., 1995, MNRAS, 275, 381
  \bibitem[\protect\citename{Nagase }1989]{nag89}
    Nagase~F., 1989, PASJ, 41, 1
  \bibitem[\protect\citename{Porter }1996]{por96}
    Porter~J.~M., 1996, MNRAS, 280, L31
  \bibitem[\protect\citename{Schmidt-Kaler }1996]{smk82}
    Schmidt-Kaler~Th., 1982, in Scheifers~K., Voight~H.~H., eds., Landolt
    B\"ornstein New Series, Vol.\ 2b. Springer-Verlag Berlin
  \bibitem[\protect\citename{Schwarzschild }1958]{sch58}
    Schwarzschild~M., 1958, Structure and evolution of stars, Princeton 
    Univ.\ Press, Princeton
  \bibitem[\protect\citename{Smarr \& Blandford }1976]{sb76}
    Smarr~L.~L., Blandford,~R.~D., 1976, ApJ, 207, 574
  \bibitem[\protect\citename{Slettebak, Kuzma \& Collins }1976]{skc80}
    Slettebak~A., Kuzma~T.~J. \& Collins~G.~W., 1980, ApJ, 242, 171
  \bibitem[\protect\citename{Standish }1982]{sta82}
    Standish~E.~M., 1982, A\&A, 114, 297
  \bibitem[\protect\citename{Taylor \& Weisberg }1989]{tw89}
    Taylor~J.~H., Weisberg~J.~M., 1989, ApJ, 345, 434
  \bibitem[\protect\citename{Tavani \& Arons}1997]{ta97} 
    Tavani~M., Arons~J., 1997, ApJ, 477, 439
  \bibitem[\protect\citename{Waters }1986]{wat86}
    Waters~L.~B.~F.~M., 1986, A\&A, 162, 121
  \bibitem[\protect\citename{Wex }1998]{wex98} 
    Wex~N., 1998, MNRAS, In Press, (astro-ph/9706086)
\end{thebibliography}
\end{document}